# CT Super-Resolution via Zero-Shot Learning

Zhicheng Zhang, Shaode Yu, Wenjian Qin, Xiaokun Liang, Yaoqin Xie, Guohua Cao

*Abstract*—Computed Tomography (CT) is an advanced imaging technology used in many important applications. Here we present a deep-learning (DL) based CT super-resolution (SR) method that can reconstruct low-resolution (LR) sinograms into high-resolution (HR) CT images. The method synergistically combines a SR model in sinogram domain, a deblur model in image domain, and the iterative framework into a CT SR algorithm–super-resolution and deblur based iterative reconstruction (SADIR). We incorporated the CT domain knowledge into the SADIR and unrolled it into a DL network (SADIR-Net). The SADIR-Net is a zero-shot learning (ZSL) network, which can be trained and tested with a single sinogram in the test time. The SADIR was evaluated via SR CT imaging of a Catphan$^{700}$ physical phantom and a biological ham, and its performance was compared to the other state-of-the-art (SotA) DL-based methods. The results show that the zero-shot SADIR-Net can indeed provide a performance comparable to the other SotA methods for CT SR reconstruction, especially in situations where training data is limited. The SADIR method can find use in improving CT resolution beyond hardware limits or lowering requirement on CT hardware.

*Index Terms*—Computed Tomography (CT), Super Resolution (SR), Zero-Shot Learning, Iterative Reconstruction.

## I. INTRODUCTION

Computed Computed Tomography (CT) has been widely used in many applications such as cancer research [1], clinical diagnosis [2], and geosciences [3], among others. Spatial resolution as one of the most important performance metrics of CT has been increased tremendously over the last decades [4]. Although some latest CT technologies [5] can provide images at small voxel size, the inherent spatial resolution of CT is still far from ideal in many important biomedical applications such as tumor characterization [6] and atherosclerosis imaging [7]. Therefore, methods for achieving higher spatial resolution are highly sought after in the CT field.

It is well known that the inherent spatial resolution of a CT imaging system depends on hardware configurations [8] such as x-ray source focal spot size, detector pixel size, and acquisition geometry, as well as software algorithms such as raw data pre-processing, tomographic image reconstruction, and post-reconstruction image processing. CT super-resolution (SR) imaging is a class of techniques that can enhance the spatial resolution of a CT imaging system. Consequently, there are two categories of methods for CT SR: hardware-driven and software-driven. For hardware-driven methods, the key is to improve the hardware performance of a CT scanner with smaller X-ray source focal spot size, smaller X-ray detector pixel size, or more favorable acquisition geometry. It is easy for one to see that the hardware-based methods will inevitably increase the cost and complexity of the CT system, as well as the X-ray radiation dose if signal-to-noise (SNR) is to be maintained. The software-based methods can achieve CT SR without hardware modifications, and thus are more attractive. Furthermore, due to the well-known ALARA (as low as reasonably achievable) principle [9], software-based CT SR methods are particularly desirable for biomedical imaging.

There are two types of software-based methods for improving CT image resolution. The first type of methods can improve CT image resolution *indirectly* via artifact corrections [10]. Those indirect-type methods *cannot* improve spatial resolution of a CT scanner beyond the detector sampling limit at the object, which is determined by the detector pixel size and geometry magnification factor. The second type of methods, in contrast, can improve CT image resolution *directly*, and can achieve a CT spatial resolution beyond the detector sampling limit. The CT SR methods discussed in this paper belong to the direct-type CT SR methods. These direct-type methods can potentially enable a low-end CT scanner with a low-resolution (LR) detector to provide high-resolution (HR) CT images that are used to be available only from a high-end CT scanner with a HR detector, or can enhance a high-end CT scanner (*e.g.* Nano-CT) with a HR detector to provide an even higher CT resolution.

The direct-type CT SR methods can be further grouped into the following two sub-types: model-based SR methods, and learning-based SR methods. The essence of model-based methods is to develop more accurate physical models so that HR CT images can be generated from LR CT images or LR sinograms [11]. Model-based SR methods can be either sinogram-based, such as sinogram rebinning [12] and frequency boosting [13], or imaging-chain based [14]. The imaging-chain based algorithms fully consider the CT image formation process and explicitly model the defects in the imaging chain. Noticeably, with the emergence of compressed sensing (CS) [15], a variety of model-based iterative reconstruction algorithms with some effective priors, such as sparsity in gradient domain [16] and low rank [17], have been developed. For instance, Thibault *et al.* developed a

This work was supported in part by the National Natural Science Foundation of China (62001464,61871374), Dr. Guohua Cao's CAREER award from the US National Science Foundation (CBET 1351936) and his Commonwealth Research Commercialization Fund (MF17-022-LS), and by grants from Shenzhen matching project (GJHS20170314155751703), National Key Research and Develop Program of China (2016YFC0105102), Shenzhen Basic Research Program (JCYJ20170413162354654). (Corresponding authors: Yaoqin Xie; Guohua Cao.)

Z. Zhang and X. Liang are with Department of Radiation Oncology, Stanford University, Palo Alto, CA 94306, USA; (e-mail: zzc623@stanford.edu; xiaokun.leung@gmail.com).

S. Yu is with College of Information and Communication Engineering, Communication University of China, Beijing 100024, China; (e-mail: yushaodecuc@cuc.edu.cn).

W. Qin and Y. Xie are with Shenzhen Institutes of Advanced Technology, Chinese Academy of Sciences, Shenzhen, Guangdong 518055, China; (e-mail: {wj.qin, yq.xie}@siat.ac.cn).

G. Cao is with Virginia Polytechnic Institute & State University, Blacksburg, Virginia 24061, USA; (e-mail: ghcao@vt.edu).

model-based iterative reconstruction algorithm that incorporated a well-designed statistical noise model, an accurate forward model, a more realistic detector response model, and a voxel model that takes account of non-infinitesimal X-ray source focal spot size and discrete X-ray detector with under-sampling rate [18]. Recently, as the amount of available data increases, learning-based (before *deep learning*) SR methods have undergone extensive development. In learning-based SR methods, sparse representation such as dictionary learning [19] was employed to incorporate more low-level information from the abundant data. For instance, Zhang *et al.* [20] proposed a patch-based SR algorithm that made use of information from all-phase CT images to estimate the missing structure information and obtained SR in 4D CT images of the lung.

Lately, deep learning (DL) [21] is playing an increasingly important role in many aspects of medical imaging, including image segmentation [22], image quality assessment [23], nodule detection [24], CT reconstruction [25-26], and CT SR [27]. With a well-designed neural network, DL-based methods can extract abstract features from training data, thereby establishing a nonlinear mapping relation between measured data and ground-truth reference. For instance, a simple 3-layer convolutional neural network (CNN) [28] was first proposed to estimate HR images directly from LR images [27]. Later, CNNs were extended with more advanced operators such as residual learning, gradient clipping [29-30], and residual dense block [31], to increase the depth, expressivity, as well as computability of the neural networks. The success of DL lies in the availability of a large amount of training data, but this dependence also restricts the application of DL in some scenarios where a large amount of training data is not easily available. Currently, progress is made on designing DL models that can be trained with a small amount of training data [32-35], or even without training data (*e.g.* Zero-shot learning, ZSL [36]).

Inspired by the universal learning ability of DL models [37], and motivated in part by our difficulty in obtaining a large amount of training CT data during the early phase of Covid-19 pandemic, here we present a super-resolution and deblur based iterative reconstruction algorithm (SADIR), for the purpose of reconstructing LR sinograms into HR CT images without the need for a large amount of training data. SADIR synergistically combines a SR model in *sinogram* domain, a deblur model in *image* domain, and the iterative CT reconstruction framework. In this paper, the SADIR algorithm was evaluated by CT SR imaging of physical objects in a realistic CT system with two detector resolutions. We used the SADIR algorithm to reconstruct HR CT images from LR sinograms acquired from a LR detector, and then compared the SADIR-reconstructed HR CT images to reference HR CT images reconstructed from HR sinograms acquired from a HR detector in the same CT system. In the SADIR algorithm, the sinogram-based SR model and the image-based deblur model are linked by the relations between the sinograms and the corresponding CT images in CT SR imaging with two detector resolutions. We unrolled the SADIR algorithm into a cascaded neural network (SADIR-Net). Furthermore, we designed the SADIR-Net so that it can be trained in a self-supervised fashion using the ZSL strategy. The SADIR-Net can be trained and tested on the same LR sinogram.

After the training, the SADIR-Net can reconstruct the same LR sinogram into a HR CT image. No other training data are required. This means that SADIR-Net can be trained and tested in the test time.

The contributions of this study are twofold. First, the SADIR-Net is a *hybrid* DL method for CT SR imaging that combines the DL techniques with CT domain knowledge. SADIR-Net models after some important image formation processes in the CT super-resolution problem, including the image blurring in both the sinogram and image domains, and relations between the sinograms and the corresponding CT images in CT SR imaging with two detector resolutions. SADIR-Net learns only the relevant blur kernels of the blurring models, and the weighting factors and penalty functions of the regularization terms in the optimization solution for the CT SR problem (see details in Section V. Method). Second, SADIR-Net is a *zero-shot* method for CT SR imaging. Zero-shot neural networks are self-learning (or unsupervised-learning) neural networks. They do not require any reference data as ground truth during network training. In the case of SADIR-Net, all it needs is a single LR sinogram, which is used for both training and testing.

II. METHOD

A. *Preliminary.*

A common task in image processing is to restore or reconstruct high-quality images from measured low-quality data. This image restoration problem can be generalized as the following optimization problem:

$$\underset{X}{argmin}||AX-Y||_2^2+\lambda g(X) \qquad (1),$$

where $X$ is the high-quality image to be obtained, $Y$ is the measured low-quality data, and A denotes some imaging system matrix. $||\cdot||_2^2$ represents the $l_2$ norm. $||AX-Y||_2^2$ is the fidelity term, $g(X)$ is the regularization term, and $\lambda$ is a scalar hyper-parameter to balance between the fidelity term and regularization term. The regularization term $g(X)$ can have different expression for different prior knowledge. For example, the sparseness of image gradient leads to a total variation term [16], the rank of image results in a low-rank regularization term [38]. Other forms of regularization include edge-preserving regularization [39], *etc*.

For the SR task in sinogram domain, it can be considered as a special kind of the image restoration task. Therefore, a HR sinogram, $Y_H$, can be restored from a LR sinogram, $Y_L$, according to the following *SR model* [40]:

$$\underset{Y_H}{argmin}||D_\downarrow C * Y_H - Y_L||_2^2+\lambda g(Y_H) \qquad (2),$$

where $C$ represents a blur filter in the *sinogram* domain, $D_\downarrow$ is a down-sampling operator, and * denotes a convolutional operator.

For image deblurring tasks, the key is to find blur filter [41]. Theoretically speaking, the blurred image, $X_L$, is the result of blurring the clear image, $X_H$, by the blur filter. Therefore, the *deblur model* can be written as follows:

$$\underset{X_H}{argmin}||B*X_H-X_L||_2^2+\lambda g(X_H) \qquad (3),$$

where $B$ represents the blur filter in the *image* domain.

B. *Problem Formulation.*

Assuming two CT imaging devices with the same hardware compositions except for the X-Ray detectors (for simplicity, here we consider line detectors only), one device has a low-quality X-Ray detector with a given number of detector elements at a given element size, while the other has a high-quality X-Ray detector with twice the number of detector elements at half the element size. Therefore, both CT imaging devices have the same field-of-view (FOV) with the former giving LR sinograms while the latter giving HR sinograms (with 2X higher resolution in this case). Obviously, after reconstruction the LR sinograms will lead to LR CT images while the HR sinograms will lead to HR CT images. A working hypothesis is that the imaging process from the lower-resolution to low-resolution is the same as that from the low-resolution to high-resolution. The main objective of this work is to obtain HR CT images from LR sinograms (*i.e.* CT super-resolution) through developing a DL-based CT SR reconstruction algorithm.

The above described CT SR problem deals with the following four groups of data: LR sinogram $Y_L$, HR sinogram $Y_H$, HR CT image $X_H$, and LR CT image $X_L$. Figure 1 shows the relations between the sinograms, $Y_L$ and $Y_H$, and the corresponding CT images, $X_L$ and $X_H$. The imaging model and the corresponding relations can be expressed mathematically as the following:

$$Y_H = A_H X_H$$
$$Y_L = A_L X_L + n$$
$$Y_L = D_\downarrow C*Y_H + n \quad (1).$$
$$X_L = B*X_H$$

Here, $n$ represents noise in the LR sinogram, $A_H$ is the high-resolution projection matrix and $A_L$ demonstrates the low-resolution projection matrix.

The SADIR synergistically combines the SR model in sinogram domain as shown in equation (2) and the deblur model in image domain as shown in equation (3). Considering the imaging model as expressed in equations (1) - (4), the SADIR model can be represented as the following equation (5).

$$\underset{X_H}{argmin} \frac{\tau_1}{2}||D_\downarrow C*Y_H - Y_L||_2^2 + \frac{\tau_2}{2}||B*X_H - X_L||_2^2 + R(X_H)$$
$$\text{s.t. } X_L = F(Y_L), Y_H = A_H X_H, X_H > 0 \quad (5),$$

where $\tau_1$ and $\tau_2$ are two weighting hyper-parameters for the fidelity terms, and $F$ is the FBP operation. In equation (5), we minimize not only the fidelity term from the SR model in equation (2) but also the fidelity term from the deblur model in equation (3), and the regularization term on $X_H$ as well.

To increase the expressivity of the SADIR model on image priors, the FOE model [42] is introduced to the regularization term in equation (5), so that the FOE model can be expressed as $R(X_H) = \sum_{k=1}^{K} R_k(G_k * X_H)$, where $K$ is the number of regularizing operators, $R_k$ is the $k^{th}$ non-linear penalty function, and $G_k$ is the $k^{th}$ linear filter of size $n_G \times n_G$. The FOE model allows both the filters and penalty functions to be learned from training data. It has been reported that replacing a large-size convolutional kernel with several small-size convolutional kernels cascaded together can maintain or even expand the receptive field of convolution while reduce the number of parameters in the network [43]. Inspired by this finding, all the convolution kernels in equation (5) were replaced by three convolutional kernels with small size. Therefore, equation (5) can be represented as follows:

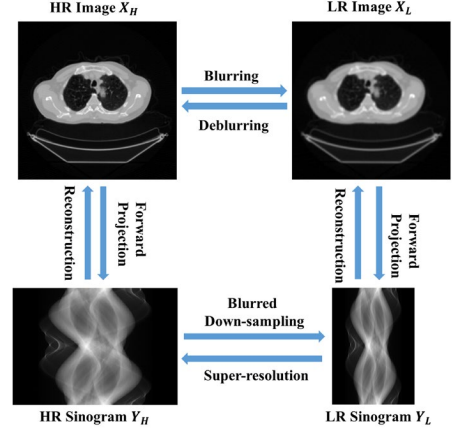

Figure 1. Relations between the sinograms and the corresponding CT images in CT SR imaging with two detector resolutions: LR sinogram $Y_L$, HR sinogram $Y_H$, HR CT image $X_H$, and LR CT image $X_L$. The HR sinogram is twice the width of the LR sinogram because the HR sinogram is from a HR detector with twice the number of detector elements at half the element size.

$$\underset{X_H}{argmin} \frac{\tau_1}{2}||D_\downarrow C_3*C_2*C_1*Y_H - Y_L||_2^2 + \frac{\tau_2}{2}||B_3*B_2*B_1*X_H - X_L||_2^2$$
$$+ \sum_{k=1}^{K} R_k(G_{k3}*G_{k2}*G_{k1}*X_H)$$
$$\text{s.t. } X_L = F(Y_L), Y_H = A_H X_H, X_H > 0 \quad (6),$$

To solve the optimization problem in equation (6), a gradient descent method was used: $X_H^{t+1} = X_H^t - \lambda_3 \nabla grad$, where $\lambda_3$ is the step size in the gradient descent method and $\nabla grad$ is the gradient form of equation (6). This leads to the following equation:

$$X_H^{t+1} = X_H^t - \{\lambda_1 F\overline{C_1}*\overline{C_2}*\overline{C_3}*D_\uparrow(D_\downarrow C_3*C_2*C_1*A_H X_H^t - Y_L)$$
$$+ \lambda_2 \overline{B_1}*\overline{B_2}*\overline{B_3}*(B_3*B_2*B_1*X_H^t - F(Y_L))$$
$$+ \lambda_3 \sum_{k=1}^{K} \overline{G_{k1}}*\overline{G_{k2}}*\overline{G_{k3}}*\varphi_k(G_{k3}*G_{k2}*G_{k1}*X_H^t)\} \quad (7),$$

Here, $\overline{C}, \overline{B}$ and $\overline{G}$ can be obtained via rotating $C, B$ and $G$ by 180 degrees [33]; $D_\uparrow$ is the up-sampling operator; $\varphi_k(x)$ is the derivative of $R_k(x)$, $\varphi_k(x) = \partial R_k(x)/\partial x$; and $F$ represents the FBP operation. FBP operation rather than back-projection was used in order to speed up the convergence of the iterative process [44].

In this work, the down-sampling operator $D_\downarrow$ was implemented by sampling at an interval of 2, and up-sampling operator $D_\uparrow$ was carried out via bilinear interpolation method. Additionally, to parameterize the penalty functions in the regularization term, the Gaussian Mixture Model (GMM) [45] was adopted. The adaption of GMM leads to the following equation (8), for the function φ(x):

$$\varphi(x) = \sum_{n=1}^{N} \gamma_n \exp\left(-\frac{(x-\mu_n)^2}{2\delta_n}\right) \quad (8),$$

where $\gamma_n$ is the weighting factor, $\mu_n$ denotes the expectation value, $\delta_n$ is the precision value, and $N$ is the number of Gaussian functions in the GMM.

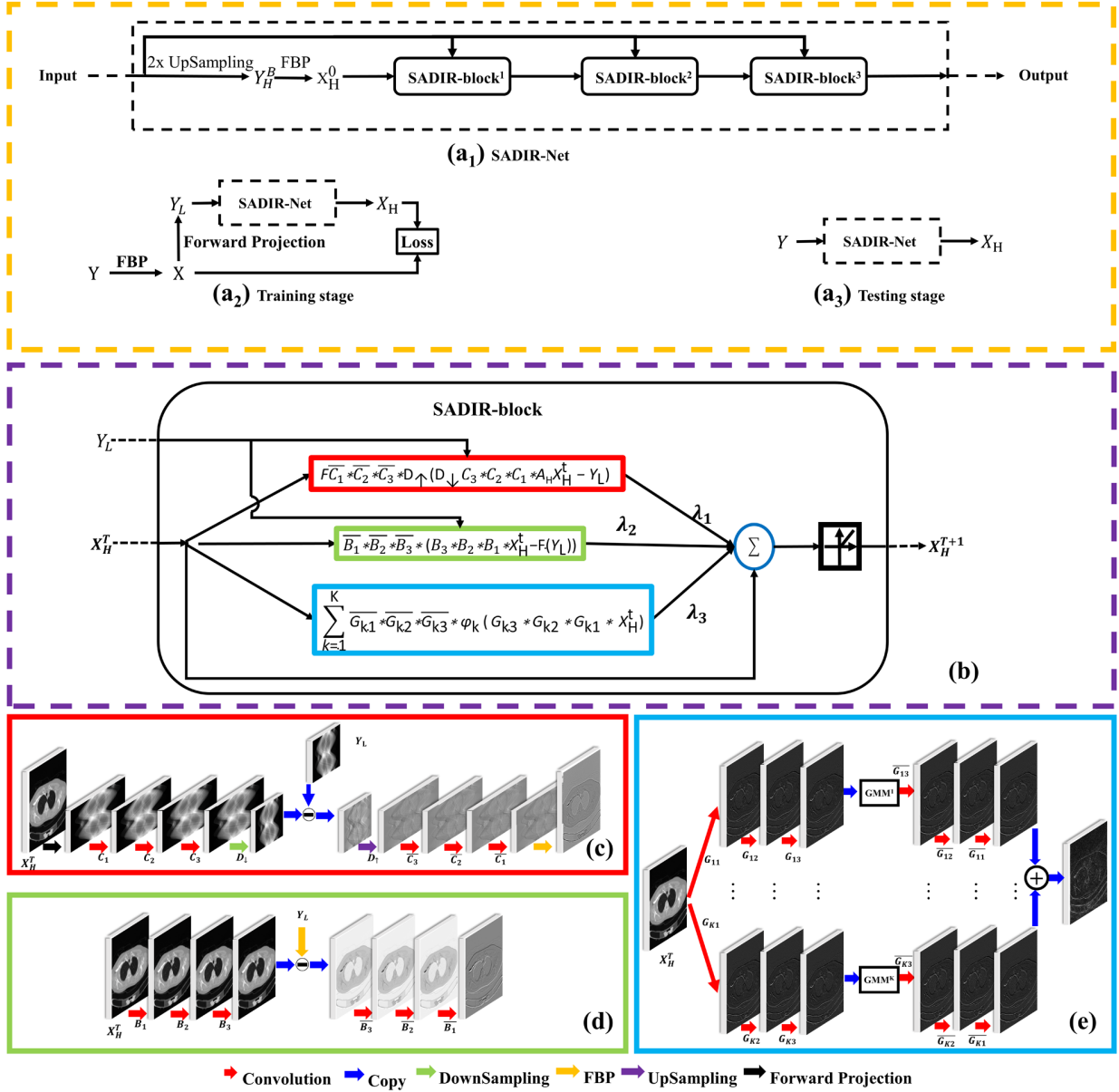

Figure 2. The architecture of the SADIR-Net. (a$_1$). A high-level illustration of the proposed SADIR-Net. Its input is a LR sinogram and its output is a HR CT image. (a$_2$) The training stage, during which a LR sinogram $Y_L$ is simulated from the acquired sinogram $Y$ to train the SADIR-Net. (a$_3$) The testing stage, during which the acquired sinogram $Y$ is reconstructed by the trained SADIR-Net into a HR CT image. (b) A detailed view of the SADIR-block. (c-d) CNN implementations of the composition terms in the SADIR-block. Each composition term in (b) and its CNN implementation are marked with a rectangle box with same color.

### C. SADIR-Net Architecture.

In this work, we unrolled the optimization algorithm in equation (7) and converted the SADIR model into a CNN-based cascaded network (SADIR-Net). The SADIR-Net is inherently iterative, with the number of iterations empirically set at 3 in this study, as shown in Figure 2(a$_1$). At each iteration is a SADIR-block. During the network training, the initial input for SADIR-block is generated from the acquired sinogram, $Y$. From $Y$, a corresponding CT image, $X$, can be reconstructed with the FBP algorithm. From $X$, a LR sinogram, $Y_L$, can be forward-projected according to reference [26]. In this study, the objective is to achieve CT SR by a factor of 2. Consequently, if the starting sinogram, $Y$, has a dimension of $i \times 2j$ ($i$ is the number of projections and $2j$ is the number of detector pixels in the line detector), the LR sinogram, $Y_L$, would have a dimension of $i \times j$. From the LR sinogram of a dimension $i \times j$, $Y_L$, a HR sinogram of a dimension $i \times 2j$, $Y_H^B$, can be obtained with 2X up-sampling via the bilinear interpolation method. From the up-sampled sinogram, $Y_H^B$, the initial HR CT image, $X_H^0$, can be reconstructed with the FBP algorithm. The initial HR CT image, $X_H^0$, is then used as the initial input to the SADIR-block iteration. After 3 SADIR-block iterations, the HR CT image output from the final SADIR-block, $X_H$, is compared to the original CT image, $X$, and their difference is factored into the loss function for the SADIR-Net. In the training stage shown in Figure 2($a_2$), a LR sinogram $Y_L$ is simulated from the acquired sinogram $Y$ and this LR sinogram is used to train the SADIR-Net. In the testing stage shown in Figure 2($a_3$), we apply the trained SADIR-Net to reconstruct the acquired sinogram $Y$ into the HR CT image. Therefore, the SADIR-Net can be trained and tested on a single sinogram, and hence is a zero-shot learning network.

The architecture for each SADIR block is shown in Figure 2(b). The input to a SADIR-block is the output from the previous SADIR block, as well as the LR sinogram (i.e. $Y_L$). Consistent with equation (7), the output of the $(t+1)^{th}$ SADIR-block, $X_H^{t+1}$, is a linearly weighted sum of the following four parts: the output of the $t^{th}$ SADIR-block ($X_H^t$), the differential form of the fidelity term in the SR model ($\overline{C_1}*\overline{C_2}*\overline{C_3}*D_\uparrow (D_\downarrow C_3*C_2*C_1*A_H X_H^t - Y_L)$, marked by the red rectangle in Figure 2(b)), the differential form of the fidelity term in the deblur model ($\overline{B_1}*\overline{B_2}*\overline{B_3}* (B_3*B_2*B_1*X_H^t - F(Y_L))$, marked by the green rectangle in Figure 2(b)), and the differential form of the regularization term ($\sum_{k=1}^{K} \overline{G_{k1}}*\overline{G_{k2}}*\overline{G_{k3}}* \varphi_k ( G_{k3} * G_{k2} * G_{k1} * X_H^t)$, marked by the blue rectangle in Figure 2(b)).

Figure 2(c) shows the external representation of the differential form of the fidelity term in the SR model. The neural network in Figure 2(c) literally follows the formula ($F\overline{C_1}*\overline{C_2}*\overline{C_3}*D_\uparrow (D_\downarrow C_3*C_2*C_1*A_H X_H^t - Y_L)$. First, the initial HR CT image, $X_H^t$, is forward projected into the projection domain. Then, the projection result (i.e. $A_H X_H^t$) is convoluted by the three convolution filters (i.e. $C_3, C_2, C_1$), followed by a down-sampling filter, $D_\downarrow$. The result is then subtracted by the LR sinogram, $Y_L$. The subtraction result is then up-sampled by the up-sampling filter, $D_\uparrow$, followed by the three corresponding convolution filters (i.e. $\overline{C_1}, \overline{C_2}, \overline{C_3}$). Finally, result after the convolutions is FBP-reconstructed into the output of the network, $X_H^{t+1}$. Since the SR task is carried out in sinogram domain, all the filters in this network have a dimension of $1 \times n_c$ ($n_c = 3$ in this work).

Similarly, Figure 2(d) shows the external representation of the differential form of the fidelity term in the deblur model ($\overline{B_1}*\overline{B_2}*\overline{B_3}* (B_3*B_2*B_1*X_H^t - F(Y_L))$, and Figure 2(e) shows the external representation of the differential form of the FOE model ($\sum_{k=1}^{K} \overline{G_{k1}}*\overline{G_{k2}}*\overline{G_{k3}}* \varphi_k ( G_{k3} * G_{k2} * G_{k1} * X_H^t)$. For the deblur task in image domain, because the blurring can happen in two dimensions, the dimension of the blur filters $B_3, B_2, B_1$ was set at $n_b \times n_b$ ($n_b = 3$ in this work). The differential form of the FOE model has $K$ channels. In the $K^{th}$ channel $\overline{G_{k1}}*\overline{G_{k2}}*\overline{G_{k3}}* \varphi_k( G_{k3} * G_{k2} * G_{k1} * X_H^t)$, $X_H^t$ is followed by three convolutional layers, and then the result $G_{k3} * G_{k2} * G_{k1} * X_H^t$ is fed into $\varphi_k(x)$, which is the derivative of $R_k(x)$. After that, $\overline{G_{k1}}, \overline{G_{k2}}$, and $\overline{G_{k3}}$ are used to calculate the final output from the $K^{th}$ channel. The outputs from all the K channels are summed to obtain the output of the entire network in Figure 2(e). In this work, we used a linearly weighted sum of 4 Gaussian functions to approximate $\varphi_k$ (i.e. $K = 4$), and set the dimension of the convolutional kernels at $3 \times 3$.

D. *Network Training.*

To train the SADIR-Net, a joint loss function $l(\cdot)$ combining $l_2$-norm with SSIM was used to preserve CT image details [46]. Let $X$ denote the ground-truth CT image and $X_H$ denote the final HR CT image from the SADIR-Net, the joint loss function can be written as

$$l(X_H, X) = \sqrt[2]{1 + l_2(X_H, X)} \times (1 - SSIM(X_H, X)) \quad (9),$$

Here, $l_2(p,q) = \sum_{i=1}^{N}(p_i - q_i)^2$, and $SSIM(p,q) = \frac{2\bar{p}\bar{q}+\varepsilon_1}{\bar{p}^2+\bar{q}^2+\varepsilon_1} \times \frac{2\delta_{pq}+\varepsilon_2}{\delta_p^2+\delta_q^2+\varepsilon_2}$, where $N$ denotes the number of pixels in the image, $p$ and $q$ denotes the reference image and the target image, $\bar{p}$ and $\bar{q}$ is the mean intensity of image $p$ and image $q$, $\delta_p$ and $\delta_q$ is the standard deviation of image $p$ and image $q$, respectively, and $\delta_{pq}$ is the covariance of the two images $p$ and $q$. The window size in SSIM is set at 11 and the multiplication operation between $\bar{p}$ and $\bar{q}$ is element-wise. In this work, we set $\varepsilon_1 = 0.01$ and $\varepsilon_2 = 0.03$. $L$ is the range of the attenuation coefficients and was set at 4 times the linear attenuation coefficient of water at the corresponding energy level. For instance, $L$ was set at $0.082\ mm^{-1}$ when the x-ray source energy was 60 keV.

The back-propagation algorithm was used to estimate all the involved parameters, and the Adam optimizer [47] was employed to optimize the loss function. The learning rate was set at $10^{-5}$. In this work, we used a fixed number of epochs at 500 as the criteria for training convergence. Additionally, all the convolution filters were initialized with random Gaussian distributions with zero mean and 0.05 standard deviation.

The SADIR-Net is a self-supervised DL network. Furthermore, it is a "Zero-Shot" learning model because its training data and testing data are derived from the same sinogram $Y$, with the training sinogram $Y_L$ as the LR counterpart downscaled from the sinogram $Y$. This means the SADIR-Net can be trained using the testing sinogram in the test time. After the training, the same sinogram can be reconstructed by the trained SADIR-Net to obtain the HR CT image.

Our codes were run on a personal workstation with Intel Core (TM) i7-4790K CPU, 16GB RAM and a GPU card (Nvidia GTX Titan Z). The total training time for one data sample was 8 minutes. After the training, the trained SADIR-Net can reconstruct an experimentally acquired sinogram, Y, into a HR CT image. The reconstruction time was 2.3 seconds per CT image.

III. RESULTS

A. *Data and Evaluation Methods*

In this work, we compared the proposed SADIR method with some other state-of-the-art (SotA) reference methods: Bicubic, SRCNN [48], VDSR [49], EDSR [50-52], ZSSR [32] and GAN-CIRCLE [53]. The classical Bicubic method is an analytical reference method. The ZSSR method is a SotA zero-shot reference method. All the other reference methods are supervised learning methods. While the supervised methods require a large amount of data for network training, the zero-shot methods do not. ZSSR can be trained and tested on a single LR CT image, and SADIR can be trained and tested on a single LR sinogram.

To train all the supervised methods, we prepared a large amount of training data. The training data were prepared as follows. A total of 47 clinical CT scans were downloaded from The Cancer Imaging Archive (TCIA) [54]. Each CT scan has hundreds of CT slice images. From each CT scan, 30 CT slice images of size 512×512 were randomly selected, to make up a total of 1410 CT images. The 1410 CT images were used as the training data for the supervised methods. For the 1410 CT images, their corresponding HR sinograms with a dimension of 360×1024 and LR sinograms with a dimension of 360×512 were generated using a sinogram-generation process described in the reference [26]. The HR sinograms were reconstructed into their corresponding HR CT images of size 1024×1024 and the LR sinograms were reconstructed into their corresponding LR CT images of size 512×512. To increase the number of training samples to better train all the supervised methods, a total of 24688 LR image patches with a size of 64×64 were used as the inputs, and a total of corresponding 24688 HR image patches with a size of 128×128 were used as the ground truths during the training of those supervised methods. After the network training, the supervised methods were applied to the FBP (i.e. filtered back-projection)-reconstructed LR CT images of a Catphan$^{700}$ phantom and a ham(see more details below), to obtain their corresponding HR CT images. The reconstructed LR and HR CT image size was 512×512 and 1024×1024, respectively.

To evaluate the performance of all the methods in real-world applications where a large number of training data is not available, a standard Catphan$^{700}$ phantom was CT imaged at two detector resolutions via detector *hardware* binning, to generate realistic sinograms and CT images. Modulation transfer function (MTF) analysis was carried out to evaluate the improvement of spatial resolution by the different methods. To further test the performance of all the methods for biomedical applications, cone-beam CT (CBCT) imaging of a ham was carried out to generate realistic sinograms and CT images under two detector resolutions via detector *hardware* binning. To compare the performance of the different methods, we carried out visual inspection and quantitative metric evaluation such as root-mean-square-error (RMSE), structure similarity index (SSIM) [55], and information fidelity criterion (IFC) [56] for the CT SR images generated by all the methods.

### B. Results from Catphan$^{700}$ Phantom

In this study, a standard quantitative phantom (*e.g.* Catphan$^{700}$) was CT scanned to generate a HR sinogram under the following acquisition parameters with a customized CBCT system: X-ray tube voltage 120 kV, X-ray tube current 1.6 mA, X-ray tube focal spot size 0.4 mm, X-ray detector with 1024 detector elements at 0.417 mm element size, source and detector distance 1500 mm, source and object distance 1000 mm, 750 projections over 360 degrees. The resulting HR sinogram has a dimension of 750×1024. By hardware-binning the X-ray detector with a factor of 2 and repeating the CT imaging experiment, a LR sinogram with a dimension of 750×512 corresponding to 512 detector elements at 0.834 mm element size was acquired. This LR sinogram was used to simulate a lower-resolution sinogram, and the lower-resolution sinogram was used to train the SADIR-Net. After the training, the SADIR-Net can reconstruct the LR sinogram into a HR CT image.

The HR sinogram was FBP-reconstructed to receive the reference HR CT image. For all the reference methods, the LR sinogram was reconstructed by FBP to produce the LR CT image, and then the resulted LR CT images was fed into the reference methods to produce their corresponding HR CT images. In Figure 3, we show the reference HR CT image in Figure 3(A), as well as the zoom-ed in views of the red rectangle region (shown in Figure 3(A)) in the CT images reconstructed by the different methods. From Figure 3(B-I), it is clear that different methods have different levels of resolution improvement. The highest line pairs that can be visually distinguished are highlighted by the white arrows in Figure 3. We can see that our SADIR method can distinguish a higher number of line pairs than the reference methods.

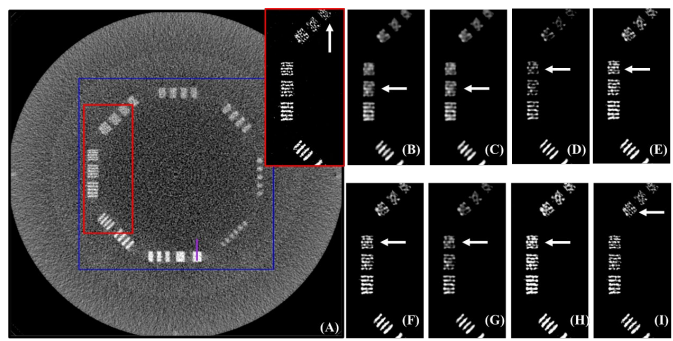

Figure 3. (A) The reference HR CT image reconstructed from the HR sinogram for the Catphan phantom. (B-I) The zoomed-in images of the red-rectangle region in (A) reconstructed by different methods: (B) LR CT; (C) Bicubic; (D) SRCNN; (E) VDSR; (F) EDSR; (G) GAN-CIRCLE; (H) ZSSR; and (I) SADIR. The display window of (A) is [0.0123, 0.0328] $mm^{-1}$. The display window of (B-I) is [0.0236, 0.0285] $mm^{-1}$.

Figure 4 shows the corresponding absolute difference images compared to the reference HR CT image, calculated from the entire phantom CT image. We can see that there is large amount of difference in the central regions of the absolute difference images from the reference methods. In contrast, we can see that the CT image reconstructed by SADIR is closest to the reference HR CT image.

**Modulation Transfer Function (MTF) Analysis.** To evaluate the improvement of spatial resolution from all the SR methods more accurately, MTF was calculated using an edge method described in the reference [57]. Figure 5 shows the MTF curves from all the methods by using the edge spread functions obtained through the purple line marked in Figure 3(A). From Figure 5, we can see that, compared to the LR MTF obtained from the LR CT image, all the SR methods can clearly improve the MTF, and all DL-based SR methods are better than the Bicubic method. Among the DL-based SR methods, SADIR is slightly better than ZSSR, VDSR and EDSR, and much better than SRCNN and GAN-CIRCLE.

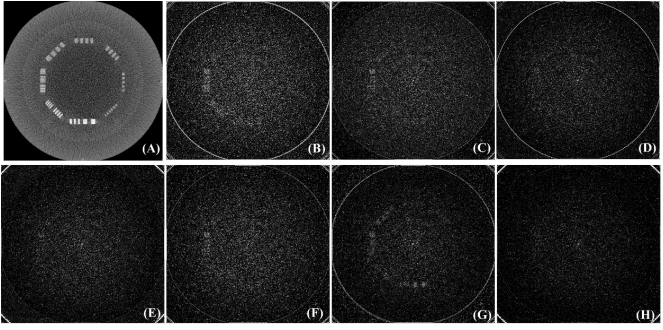

Figure 4. The corresponding absolute difference images: (A) HR CT image; (B) Bicubic; (C) SRCNN; (D) VDSR; (E) EDSR; (F) GAN-CIRCLE; (G) ZSSR and (H) SADIR. The display window of (B-H) is [0.001, 0.004] $mm^{-1}$.

In Table I, we show the quantitative results from different SR methods for the CATPHAN[700] phantom. Image quality metrics such as IFC, RMSE, and SSIM are listed. The MTF values at $MTF_{50\%}$ and $MTF_{10\%}$ obtained from their MTF curves shown in Figure 5 are also listed. From Table I, we can clearly see that SADIR gives the best image quality in terms of IFC, RMSE, SSIM, as well as the best spatial resolutions at $MTF_{50\%}$ and $MTF_{10\%}$. Quantitatively, the spatial resolutions at $MTF_{50\%}$ and $MTF_{10\%}$ from SADIR can reach 91.3% and 89.3% of the counterparts from the HR CT image that was reconstructed from the HR sinogram with FBP. Alternatively, if the LR CT image is used as the reference, SADIR can improve the spatial resolution at $MTF_{50\%}$ and $MTF_{10\%}$ by 69.2% and 69.5%, respectively.

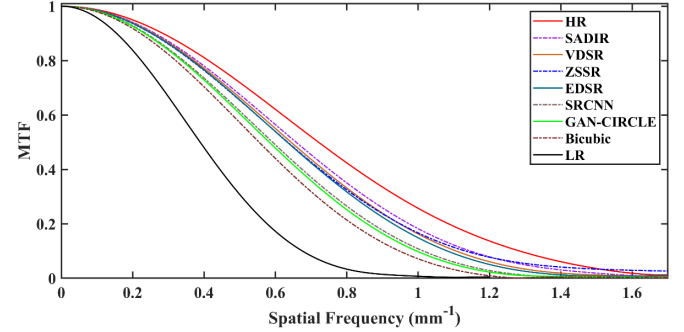

Figure 5. The MTF results from the HR reference, the LR reference, and different SR reconstruction methods.

TABLE I. QUANTITATIVE RESULTS FROM DIFFERENT METHODS FOR THE CATPHAN[700] PHANTOM USING THE REGION MARKED BY THE BLUE RECTANGLE IN FIGURE 3 (A). RED AND BLUE INDICATE THE BEST AND THE SECOND-BEST RESULTS, RESPECTIVELY.

|  | HR | LR | Bicubic | SRCNN | VDSR | EDSR | GAN-CIRCLE | ZSSR | SADIR |
|---|---|---|---|---|---|---|---|---|---|
| IFC | - | - | 1.6 | 2.009 | 2.457 | 2.623 | 1.941 | 2.326 | 3.073 |
| RMSE | - | - | 0.00169 | 0.00154 | 0.00132 | 0.00133 | 0.00147 | 0.00126 | 0.00109 |
| SSIM | - | - | 0.844 | 0.875 | 0.899 | 0.899 | 0.866 | 0.899 | 0.925 |
| $MTF_{50\%}$ | 0.721 | 0.389 | 0.554 | 0.592 | 0.644 | 0.633 | 0.583 | 0.635 | 0.658 |
| $MTF_{10\%}$ | 1.284 | 0.676 | 0.948 | 1.014 | 1.106 | 1.082 | 0.996 | 1.133 | 1.146 |

### C. Results from Ham CT Imaging

To test the performance of the SADIR method under realistic conditions, we scanned a biological object (*e.g.* a ham) using the CT scanner that was used to image the Catphan[700] phantom. All the scanning conditions were the same except the parameters of the detector. The ham CT imaging experiment generated a HR sinogram with a dimension of $720 \times 1024$ via operating the detector in the 1×1 binning mode with detector element size at 0.139×0.139 $mm^2$, and a LR sinogram with a dimension of $720 \times 512$ via operating the detector in the 2×2 binning mode with detector element size at 0.278×0.278 $mm^2$ at same dose levels and other CT acquisition conditions. The HR sinogram was FBP-reconstructed to obtain the reference HR CT image. The LR sinogram was reconstructed by FBP to produce the LR CT image, which was then fed into all the reference CT SR methods to obtain their HR CT images. For SADIR, we first used the SADIR-Net which was trained with the LR sinogram from the Catphan[700] phantom (*i.e.* Catphan-trained SADIR), to reconstruct the LR sinogram of the ham into the corresponding HR CT image. Because SADIR is a unique method that can model the image formation processes in the physical CT system, its performance will obviously benefit from training and testing on the data from the same object. Therefore, for SADIR we carried out another experiment by training SADIR with the LR sinogram of the ham (*i.e.* Ham-trained SADIR), and then reconstructing the LR sinogram of the ham into another HR CT image. The reconstructed LR and HR CT image size was 512×512 and 1024×1024, respectively.

After reconstruction, a zoomed-in region that corresponds to a bone structure within the ham is shown in Figure 6. The ability of different methods to preserve the complex and heterogeneous features inside the bone can serve as a good indicator for the resolution improvement of the different methods. Through visual comparison, we can see that the performance of the Catphan-trained SADIR is similar to the ones from ZSSR, VDSR and EDSR, and slightly better than the ones from SRCNN and GAN-CIRCLE. Furthermore, the image in Figure 6(J) from the ham-trained SADIR appears to have a clearer texture than the one in Figure 6(I) from the Catphan-trained SADIR. This indicates that SADIR indeed can benefit from training and testing on the data from the same object. At a closer inspection, the interosseous spatium as well as other internal textures of the bone are more visible in the CT images from SADIR, ZSSR, VDSR, and EDSR while they appear vague in the CT images from SRCNN and GAN-CIRCLE. Additionally, Figure 7 shows the absolute difference images for the same bone region within the ham. From Figure 7, we can see that SADIR, ZSSR, VDSR, and

EDSR produce better resolution improvement, followed by SRCNN and GAN-CIRCLE. This observation is consistent with the above findings from Figure 3-5.

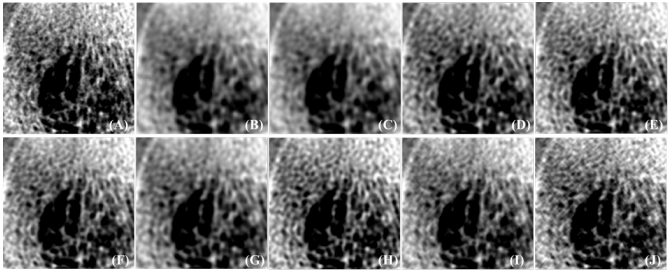

Figure 6. (A) A zoomed-in region of the reference HR CT image reconstructed from the HR sinogram of the ham. (B)-(K) The same zoomed-in region of the CT images reconstructed from different methods: (B) FBP; (C) Bicubic; (D) SRCNN; (E) VDSR; (F) EDSR; (G) GAN-CIRCLE; (H) ZSSR; (I) Catphan-trained SADIR; and (J) Ham-trained SADIR. The display window is $[0.02, 0.045]\ mm^{-1}$.

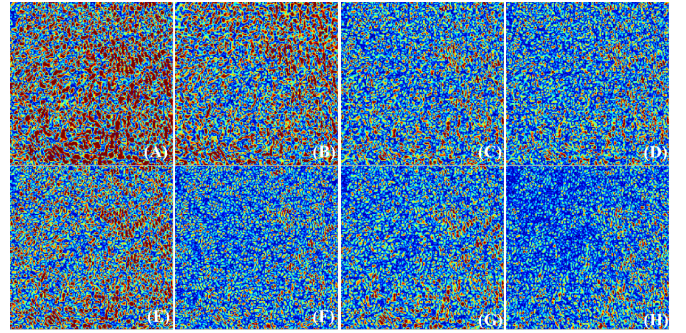

Figure 7. (A)-(F) The absolute difference images from different methods relative to the reference HR image for the zoomed-in bony region shown in Figure 6: (A) Bicubic; (B) VDSR; (C) VDSR; (D) EDSR; (E) GAN-CIRCLE; (F) ZSSR; (G) Catphan-trained SADIR; and (H) Ham-trained SADIR. Green and red color indicate small and large errors, respectively.

TABLE II. QUANTITATIVE RESULTS FROM DIFFERENT METHODS FOR THE BONE REGION OF THE HAM SHOWN IN FIGURE 6. RED AND BLUE INDICATE THE BEST AND THE SECOND-BEST RESULT, RESPECTIVELY.

|  | Bicubic | SRCNN | VDSR | EDSR | GAN-CIRCLE | ZSSR | SADIR Trained on Catphan[700] | SADIR Trained on Ham |
|---|---|---|---|---|---|---|---|---|
| IFC | 2.013 | 2.564 | 3.165 | 3.424 | 2.457 | 3.221 | 3.689 | 3.858 |
| RMSE | 0.00234 | 0.00190 | 0.00138 | 0.00129 | 0.00178 | 0.00118 | 0.00129 | 0.00107 |
| SSIM | 0.829 | 0.903 | 0.937 | 0.942 | 0.885 | 0.943 | 0.943 | 0.959 |

## IV. DISCUSSION AND CONCLUSION

In this paper, we analyzed the mathematical relations between the sinograms and the corresponding CT images in CT SR imaging with two detector resolutions. Those relations were incorporated into a new CT SR reconstruction method – SADIR. SADIR synergistically integrates the SR model in sinogram domain and the deblur model in image domain into an iterative reconstruction algorithm. Because both the SR model and the deblur model are very complex and hard to model analytically, we converted SADIR into a CNN-based cascaded network (SADIR-Net). SADIR-Net is a hybrid and self-supervised neural network, which can adaptively learn all the involved parameters via zero-shot learning. Furthermore, we employed the field of expert (FOE) model for the regularization term in the SADIR algorithm, to increase the expressivity of SADIR-Net. Through CT imaging of a Catphan[700] phantom and a ham, we demonstrated that SADIR can effectively reconstruct realistic LR sinograms into high-quality HR CT images, and SADIR clearly outperformed the other SotA DL-based methods such as ZSSR, SRCNN, VDSR, EDSR, and GAN-CIRCLE.

To the best of our knowledge, SADIR-Net is a new, *hybrid* DL method for CT SR imaging. SADIR-Net incorporates the CT domain knowledge into the DL model, which is different from the traditional, *DL-only* methods such as SRCNN, VDSR, EDSR, and GAN-CIRCLE. SADIR-Net was designed via unrolling an optimization solution to the CT SR reconstruction problem. This way of designing SADIR-Net has the following two advantages. First, the unrolled SADIR-Net is more interpretable than a traditional "black-box" type DL model such as U-Net [58]. The SADIR-Net is unrolled from equation (7), which is our mathematically derived solution to the general CT SR reconstruction problem. Designing SADIR-Net in this fashion ensures that the CT domain knowledge (such as forward projection and back projection between sinograms and images) is incorporated into the DL model, and the DL model only needs to learn the blur filters in sinogram and image domains as well as the other parameters in the FOE model. As a result, SADIR-Net retains much expressive power of the CT imaging model. Second, SADIR-Net has much smaller number of network parameters than the traditional DL networks. In this study, the number of parameters in SADIR is 2.6% of SRCNN, 0.1% of ZSSR, 0.03% of VDSR, 0.008% of EDSR, and 0.004% of GAN-CIRCLE.

To the best of our knowledge, SADIR-Net is a zero-shot method for CT SR imaging. Traditional neural networks for CT SR imaging are supervised networks. Supervised networks require large amount of labeled LR-HR image pairs, with the HR images acting as the corresponding ground truths during network training. During the training, a supervised network learns a *"forced"* correlation between the LR-HR image pairs. The correlation learned through this approach is more like a *"black box"*, which typically has low explainability and weak robustness. Because different CT scanners and different CT imaging protocols often have different data acquisition configurations, to obtain the optimum CT SR performance, a supervised network trained for one CT scanner or CT imaging protocol will require re-training of the network with a new set of large training data collected from the new configuration.

Compared to supervised networks, zero-shot networks such as SADIR-Net have better explainability, more robustness, and easier trainability. As a result, zero-shot methods typically have a much smaller number of network parameters, which leads to the following two advantages of SADIR in CT SR reconstruction. First, SADIR-Net can relieve us from the burden of preparing a large amount of training data. SADIR-Net does not need any training data, nor does it require HR CT image to act as ground truth during network training. Second, SADIR-Net demands less computing cost during network training. While supervised networks could take hours or even days to train on a high-performance computer, SADIR-Net can be trained on a personal workstation for just 8 minutes.

Of the two zero-shot methods investigated in this study (*i.e.* ZSSR and SADIR), it is worth to point out that ZSSR is an image-domain based method while SADIR is a cross-domain (both sinogram and image domains) based method. ZSSR was originally suggested for the SR tasks in natural images[32], and here it was applied to generate HR CT images from their corresponding LR CT images. ZSSR is a simple multi-layer CNN, which treats the SR task as a "black box" and does not incorporate any domain knowledge. In contrast, SADIR-Net models after the imaging processes in the CT SR problem, and incorporates the corresponding domain knowledge.

Although promising, SADIR has following two limitations. First, SADIR works best for CT SR imaging in a CT system *only* when a large amount of training data from the *same* CT system is not available. SADIR incorporates some CT domain knowledge (such as the relations between sinograms and CT images), and SADIR-Net tries its best to learn some hard-to-model CT SR imaging parameters (such as the blur filters) of the CT system from a single LR sinogram. Therefore, both the size of SADIR-Net and the number of the parameters in SADIR-Net are relatively small compared to those supervised networks (*i.e.* SRCNN, VDSR, EDSR, and GAN-CIRCLE). Consequently, SADIR-Net cannot perform as well as those supervised networks for CT SR imaging in a CT system when a large amount of training data from the same CT system is available. Those supervised networks can learn from a large amount of training data collected from the same CT system during network training, thus the well-trained supervised networks can better approximate the CT SR imaging process in the CT system and have the better performance than the unsupervised methods (*i.e.* SADIR and ZSSR). This has been demonstrated by the additional experiments shown in the supplementary material. Nevertheless, as shown by all the results in this study, one of the main advantages of SADIR-Net is that it can easily adapt on a new dataset from a new machine, while it is not feasible for the supervised learning methods to collect enough data to train a model for every new case. Second, SADIR has a tradeoff between performance and speed if it is to be used in real-time. As shown in Figure 6, Figure 7 as well as Table II, the ham-trained SADIR-Net has better performance than the Catphan-trained SADIR-Net when testing on the LR sinogram from the ham. This indicates that SADIR-Net will have the best performance in CT SR imaging of an object if it is trained on the sinogram data from the *same* object. Since the training of SADIR-Net take times (8 min for one sinogram), SADIR cannot achieve its optimum performance in real-time.

In conclusion, a super-resolution and deblur based iterative reconstruction algorithm (SADIR) has been developed for CT SR beyond the sampling limit. The SADIR algorithm has been implemented as a zero-shot learning network (SADIR-Net). SADIR-Net synergistically incorporates the CT domain knowledge into the DL model. CT SR imaging experiments with a physical phantom and a biological object have demonstrated that SADIR can provide a CT SR reconstruction performance comparable to the other SotA methods. SADIR has great potential for obtaining HR CT images from CT scanners with LR CT detectors or achieving even higher resolution in CT machines with HR detectors. It will either lower CT hardware cost or improve spatial resolution of high-end CT such as Nano-CT, where resolution is very difficult to improve via hardware.